\documentclass[prd,amsmath,amssymb,floatfix]{revtex4}

\usepackage{graphicx}

\begin{document}

\title{Punctuated Chirality}

\author{Marcelo Gleiser}
\email{gleiser@dartmouth.edu}

\author{Joel Thorarinson}

\author{Sara Imari Walker}

\affiliation{Department of Physics and Astronomy, Dartmouth College
Hanover, NH 03755, USA}

\begin{abstract}
Most biomolecules occur in mirror, or chiral, images of each other. However, life is homochiral: proteins contain almost exclusively L-amino acids, while only D-sugars appear in RNA and DNA. The mechanism behind this fundamental asymmetry of life remains an open problem. Coupling the spatiotemporal evolution of a general autocatalytic polymerization reaction network to external environmental effects, we show through a detailed statistical analysis that high intensity and long duration events may drive achiral initial conditions towards chirality. We argue that life's homochirality resulted from sequential chiral symmetry breaking triggered by environmental events, thus extending the theory of punctuated equilibrium to the prebiotic realm. Applying our arguments to other potentially life-bearing planetary platforms, we predict that a statistically representative sampling will be racemic on average. 

\end{abstract}


\maketitle

\section{\textbf{Introduction}}

Louis Pasteur was the first to realize, in the late 1840s, that many biomolecules display a mirror asymmetry known as chirality \cite{Pasteur}. Although solutions synthesized in the laboratory are always racemic, with equal concentrations of each chiral isomer (enantiomer), the same is not true of living matter: proteins contain almost exclusively L-amino acids, while only D-sugars appear in RNA and DNA \cite{Cline}. Several mechanisms have been proposed to explain life's homochirality, although none is conclusive \cite{Fitz}. Chiral biasing effects from parity violation (PV) in the weak interactions of particle physics \cite{KN85, Bakasov} are not viable \cite{G}, while prebiotic biasing from circularly polarized UV light (CPL) originating in active star-forming regions \cite{Bailey, G} remains controversial \cite{Cataldo}.  Here, we propose that the homochirality of life may be explained by extending to prebiotic times the punctuated equilibrium hypothesis of Eldredge and Gould, whereby speciation occurred through alternating periods of stasis and intense activity prompted by external influences \cite{Eldredge}. We note that we are borrowing the concept of punctuated equilibrium with some freedom: the network of chemical reactions we describe is a non-equilibrium open system capable of exchanging energy with the environment. The periods of stasis that develop correspond to steady-states in that even though environmental influences may be negligible, chemical reactions are always occurring so as to keep the average concentrations at a constant value.

A much-debated question is whether the observed homochirality of biomolecules is a prerequisite for life's emergence or if it developed as its consequence \cite{Cohen, Bonner}. It has been argued that achiral peptide nucleic acids can form base-pairs and helical structures and thus are attractive candidates for the first carriers of genetic information \cite{Nielsen}. Considering that very specific conformations of structural entities such as $\alpha$-helices and $\beta$-sheets can only form from enantiomerically pure building blocks \cite{Cline, Fitz}, we here take the bottom-up approach and assume that the conditions for the subsequent development of complex biomolecules had to be chiral. A related question is whether chiral compounds were fed into the prebiotic soup from outer space \cite{Chyba}. The discovery of an excess of chiral organic compounds in the Murchison meteorite has supported this view \cite{Cronin, Cronin97}. Examination of isotopic distributions (in $^{15}$N/$^{13}$N) and of $\alpha$-branched amino acids of extraterrestrial origin has eliminated the possibility of contamination by Earth's biosphere, with measurements showing an excess of L-alanine of 50\% and of 30\% for L-glutamic acid \cite{Engel97}. A plausible explanation is that the cloud that originated the Solar System was subjected to CPL, such as synchrotron radiation from a neutron star. If this is indeed a viable mechanism, the same chiral bias should be prevalent throughout the Solar System but not necessarily throughout the galaxy. This should be contrasted with bias from PV, which should be the same throughout the universe. Finding an excess in D-chiral compounds in Mars \cite{Kminek} or elsewhere in the Solar System would contradict both scenarios.

We propose an alternative point of view, that environmental effects might destroy any memory of a prior chiral bias, whatever its origin. Life's chirality is interwoven with early-EarthÕs environmental history; specifically, with how the environment influenced the prebiotic soup that led to first life. 

\section{\textbf{Environmental Effects on Spatiotemporal Polymerization}}

\subsection{Polymerization Model}

We start with Sandars' polymerization model \cite{Sandars03}, a generalization of Frank's pioneering approach \cite{Frank53}, featuring autocatalysis with enantiomeric cross-inhibition. Consider a left-handed polymer $L_n$, made of $n$ left-handed monomers, $L_1$. It may grow by adding another left-handed monomer with a rate $k_s$, or be inhibited by adding a right-handed monomer $D_1$ with a rate $k_I$. (Note that we denote D-compounds by the letter ``D'' as opposed to the notation set in Sandars' work where such molecules were denoted as ``R''.) The reaction network for $n = 1, \ldots , N$, where $N$ is the maximum polymer length in the system, can be written as:

\begin{eqnarray}\label{rxnnetwork}
L_n + L_1 & \stackrel{2k_S}{\rightarrow} L_{n+1}, \nonumber \\
L_n + D_1 & \stackrel{2k_I}{\rightarrow} L_nD_1, \nonumber \\
L_1 + L_nD_1 & \stackrel{k_S}{\rightarrow} L_{n+1}D_1, \nonumber \\
D_1 + L_nD_1 & \stackrel{k_I}{\rightarrow} D_1L_nD_1, \nonumber \\
\end{eqnarray} 
supplemented by reactions for $D$-polymers by interchanging $L \rightleftharpoons D$,  and by the production rate of monomers from the substrate: $S \stackrel{k_C C_L}{\longrightarrow} L_1$;  $S \stackrel{k_C C_D}{\longrightarrow} D_1$. $C_{L (D)}$ determine the enzymatic enhancement of $L (D)$-handed monomers, usually assumed to depend on the largest polymer in the reactor pool, $C_{L(D)} = L_N(D_N)$ \cite{Sandars03}, or on a sum of all polymers \cite{WC}. Soai's group obtained the best-known illustration of this autocatalytic mechanism with enantiomeric cross-inhibition \cite{Soai}, with dimers ($N = 2$) as catalysts \cite{Blackmond04}.

A set of coupled, nonlinear ordinary differential equations for the various concentrations, $[L_1]$,  $[D_1]$, \ldots, $[L_n]$, $[D_n]$, describes the time evolution of the reaction network of eq. \ref{rxnnetwork}, supplemented by the equation for the substrate, $d[S]/dt = Q Ð (Q_L + Q_D)$, where $Q$ is the substrate's production rate, and $Q_L - Q_D = k_Cf[S](C_L - C_D)$ gives the net chiral excess in monomer production. $f$ is the enzymatic fidelity, usually set to unity to maximize chiral separation. In Fig. \ref{fig:Ncompare}, we show numerical solutions for polymerization reactions with $N = 2, 5$, and $\infty$: starting as racemates, they all evolve toward homochirality. Also included is the solution for $N = 2$ within the adiabatic approximation, where the rate of change for dimers and the substrate is assumed to be much slower than that of monomers, that is, when $k_{S,I} << k_C$ \cite{BM}. In Gleiser and Walker \cite{GW}, a detailed study of the polymerization reaction network for various values of $N$ has shown that the trends displayed in Fig. \ref{fig:Ncompare} are true even when the effects of spatial dynamics are considered, as we will do next. Gleiser and Walker also concluded that although the adiabatic approximation predicts faster approach to steady-state conditions when compared with the full $N = 2$ model, it does produce the correct asymptotic values for the various concentrations.

\subsection{Spatiotemporal Polymerization}

\begin{figure}
\centerline{\includegraphics[width=4in,height=3in]{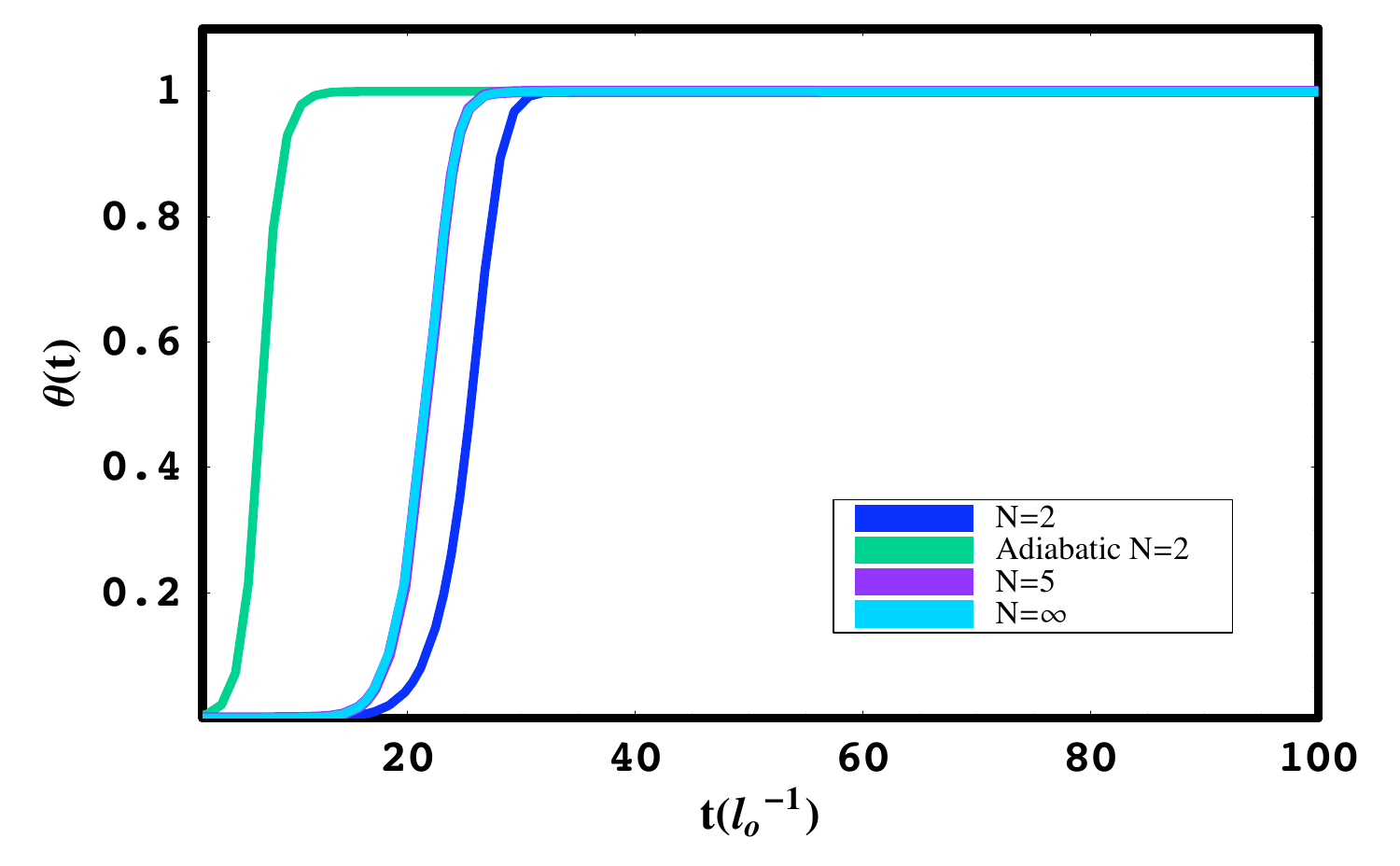}}
\caption{The time evolution of the net chirality $\theta(t) = \sum(L_n-D_n)/ \sum(L_n+D_n)$ from the reaction network of eq. \ref{rxnnetwork} for several maximum polymer lengths ($N$). All models display similar qualitative behaviour, achieving pure chirality in short time-scales. The curves for $N=5$ and $N=\infty$ overlap. $l_0 = (2k_sQ)^{1/2}$.} \label{fig:Ncompare}
\end{figure}

We now extend this treatment to include spatial dependence, establishing a reaction-diffusion network by substituting  $d/dt \rightarrow \partial / \partial t - k \nabla^2$, where $k$ is the diffusion constant \cite{BM}. In this coarse-grained approach, the number of molecules per unit volume is large enough so that the concentrations vary smoothly in space and time. The spatiotemporal evolution of the network is obtained by solving the coupled system of nonlinear PDEs for arbitrary values of $n$. Clearly, as $n$ increases, solving and statistically analyzing the coupled system of equations in two and three spatial dimensions for various parameters (see below) becomes highly CPU intensive, even for the NCSA supercomputers we used. 
 
Given that Soai's reaction \cite{Soai} -- the only illustration so far of an autocatalytic network leading to chiral purity -- featured dimers as catalysts, we focus here on the truncated system for $N = 2$ within the adiabatic approximation since, as with spatially-independent reaction networks (see Fig. \ref{fig:Ncompare}), it has similar qualitative behavior to networks with longer ($N > 2$) polymer chains \cite{GW}. The system then reduces to two coupled PDEs for the concentrations $[L_1]$ and $[D_1]$, being thus more amenable to a detailed statistical study while maintaining the key qualitative features of a larger reaction network. 

\begin{figure}[H]
\centerline{\includegraphics[width=4.5in,height=3.25in]{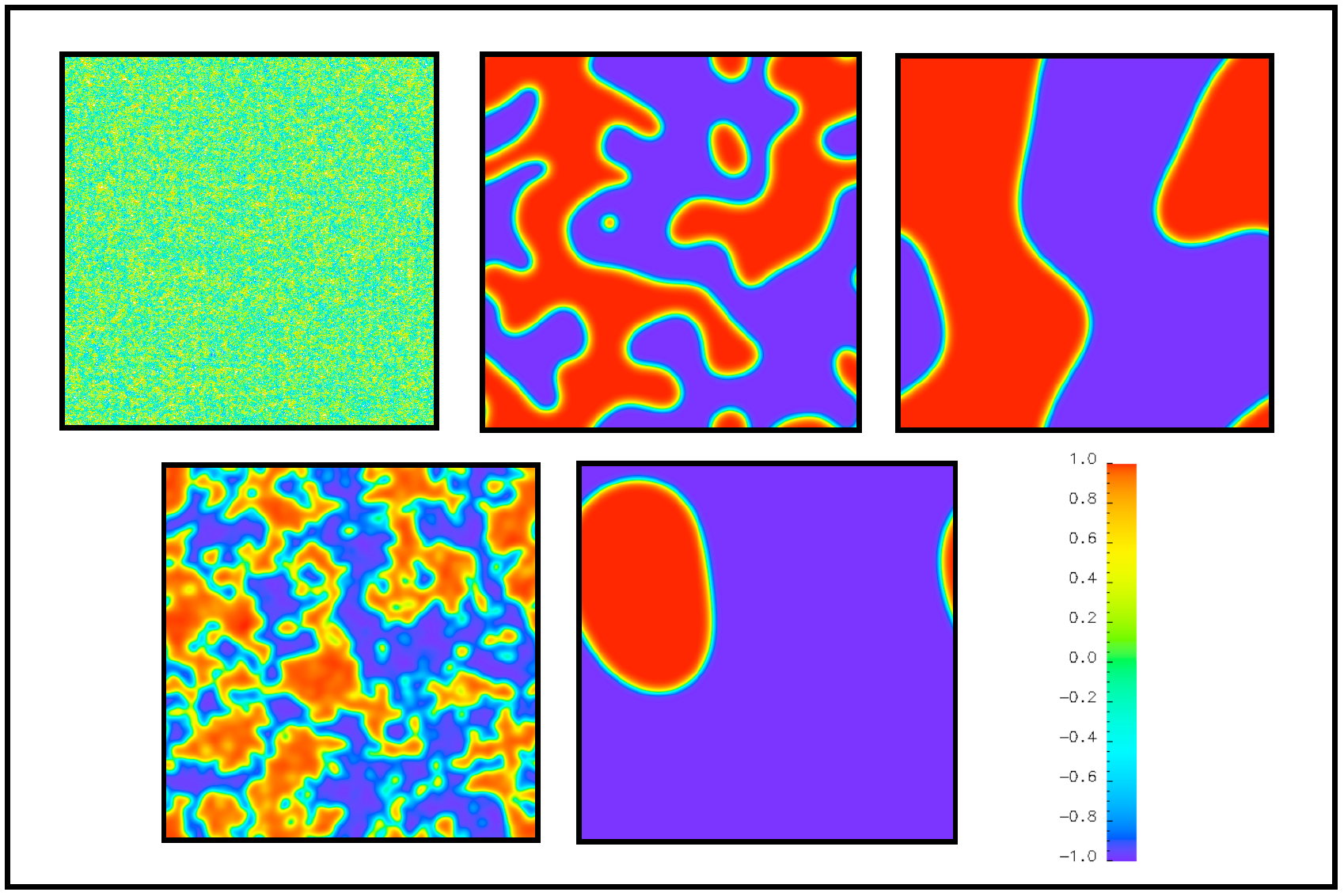}}
\caption{Evolution of $2d$ chiral domains. Red (+1 on the color bar) corresponds to the $L$-phase and blue (-1 on the color bar) corresponds to the $D$-phase. Time runs from left to right and top to bottom. Top left, the near-racemic initial conditions. Top mid and top right, evolution of the two percolating chiral domains separated by a thin domain wall. Bottom left, environmental effects break the stability of the domain wall network. Bottom right, subsequent surface-tension driven evolution leads to a enantiomerically-pure world. } \label{fig:2Dchiral}
\end{figure}

\begin{figure}[H]
\centerline{\includegraphics[width=4in,height=3in]{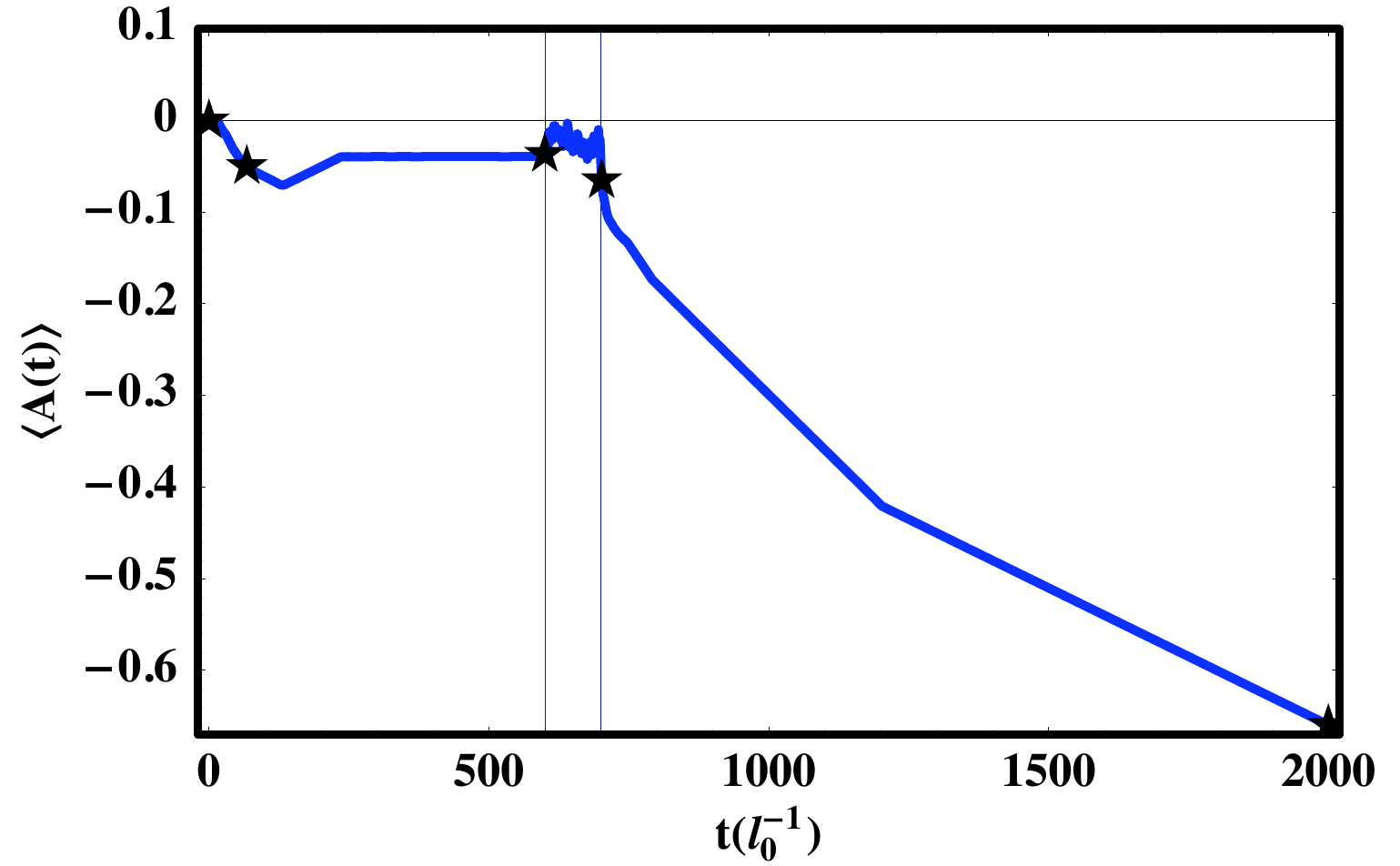}}
\caption{Time evolution of the spatially-averaged net chirality corresponding to the snapshots shown in figure \ref{fig:2Dchiral}. Stars denote times for snapshots and vertical lines mark the beginning and end of stochastic environmental  influence.} \label{fig:Aave}
\end{figure}

Introducing the variables  ${\cal S} \equiv X + Y$ and ${\cal A} \equiv X - Y$, we can rewrite the reaction network in terms of the net dimensionless chiral asymmetry ${\cal A}(t,x,y,z)$. For near-racemic initial conditions ($ |{\cal A}(0,x,y,z)| \leq10^{-4}$), the spatiotemporal evolution leads to the formation of left and right-handed percolating chiral domains separated by domain walls, as is well-known from systems in the Ising universality class (see Fig. \ref{fig:2Dchiral}). Surface tension drives the walls until their average curvature matches approximately the linear dimension of their confining volume. At this point, wall motion becomes quite slow, $d \langle {\cal A}(t) \rangle /dt \rightarrow 0$, where $\langle {\cal A}(t) \rangle$ is the spatially-averaged value of the net chiral asymmetry, and the domains coexist in near dynamical equilibrium in that the net stresses add to zero (see Fig. \ref{fig:2Dchiral}, top right). The time evolution of ${\cal A}(t)$ is shown in Fig \ref{fig:Aave}. For such model systems, it has been shown that the presence of a bias from PV or most CPL sources (even in the unlikely situation where they could be sustained unperturbed for hundreds of millions of years), would not lead to chirally-pure prebiotic conditions \cite{G}.

\subsection{Coupling Environmental Effects}

Motivated by the experimental demonstration that stirring \cite{Kondepudi, Viedma} can effectively bias chirality, and a recent numerical study for a similar $N=2$ system showing that the addition of fluid turbulence can speed up chiral evolution \cite{BM}, we proceed to examine the impact of random environmental effects in the reaction network. This is achieved in the simplest possible way via a generalized spatiotemporal Langevin equation \cite{GT}. The dynamical equations are written as:

\begin{eqnarray} \label{spatialeqns}
l_0^{-1}\left( \frac{\partial{\cal S}}{\partial t} - k \nabla^2 {\cal S} \right) &=&  1 - {\cal S}^2 + w(t, \textbf{x}),   \\ \nonumber
l_0^{-1}\left( \frac{\partial{\cal A}}{\partial t} - k \nabla^2 {\cal A} \right) &=& {\cal S}{\cal A} \left( \frac{2f}{{\cal S}^2 + {\cal A}^2} -1 \right)  +  w(t, \textbf{x}),   \\ \nonumber
\end{eqnarray}
where $l_0 \equiv (2k_SQ)^{1/2}$, and $w(\textbf{x},t)$ is a dimensionless Gaussian white noise with two-point correlation function $\langle w(\textbf{x}',t')w(\textbf{x},t) \rangle = a^2\delta(t'-t) \delta(\textbf{x}' -\textbf{x})$, where $a^2$ is a measure of the environmental influence's strength. An Ising phase diagram can be constructed showing that $\langle {\cal A} \rangle \rightarrow 0$ for $a > a_c$: chiral symmetry is restored \cite{GT}. The value of $a_c$ has been obtained numerically in two ($a_c^2 = 1.15(k/l_0^2)$cm$^2$s) and three ($a_c^2 = 0.65(k^3/l_0^5)^{1/2}$cm$^3$s) dimensions \cite{GT}. Dimensionless time, $t_0 = l_0 t $, and space, $x_0 = x(l_0/k)^{1/2}$, variables were introduced. For diffusion in water ($k = 10^{-9}$m$^2$s$^{-1}$) and nominal values $k_S = 10^{-25}$cm$^3$s$^{-1}$ and $Q = 10^{15}$ cm$^{-3}$s$^{-1}$, we obtain $l_0=\sqrt{2} \times 10^{-5}$s$^{-1}$, simulating a $2d$ ($3d$) shallow (deep) pool with linear dimensions of $l \sim 200$ ($50$) cm. For the purpose of illustration, explicit results quoted below in Section \ref{results} were computed using these values.

\subsubsection{Numerical Implementation}\label{num}

\begin{figure}
\centerline{\includegraphics[width=4in,height=3in]{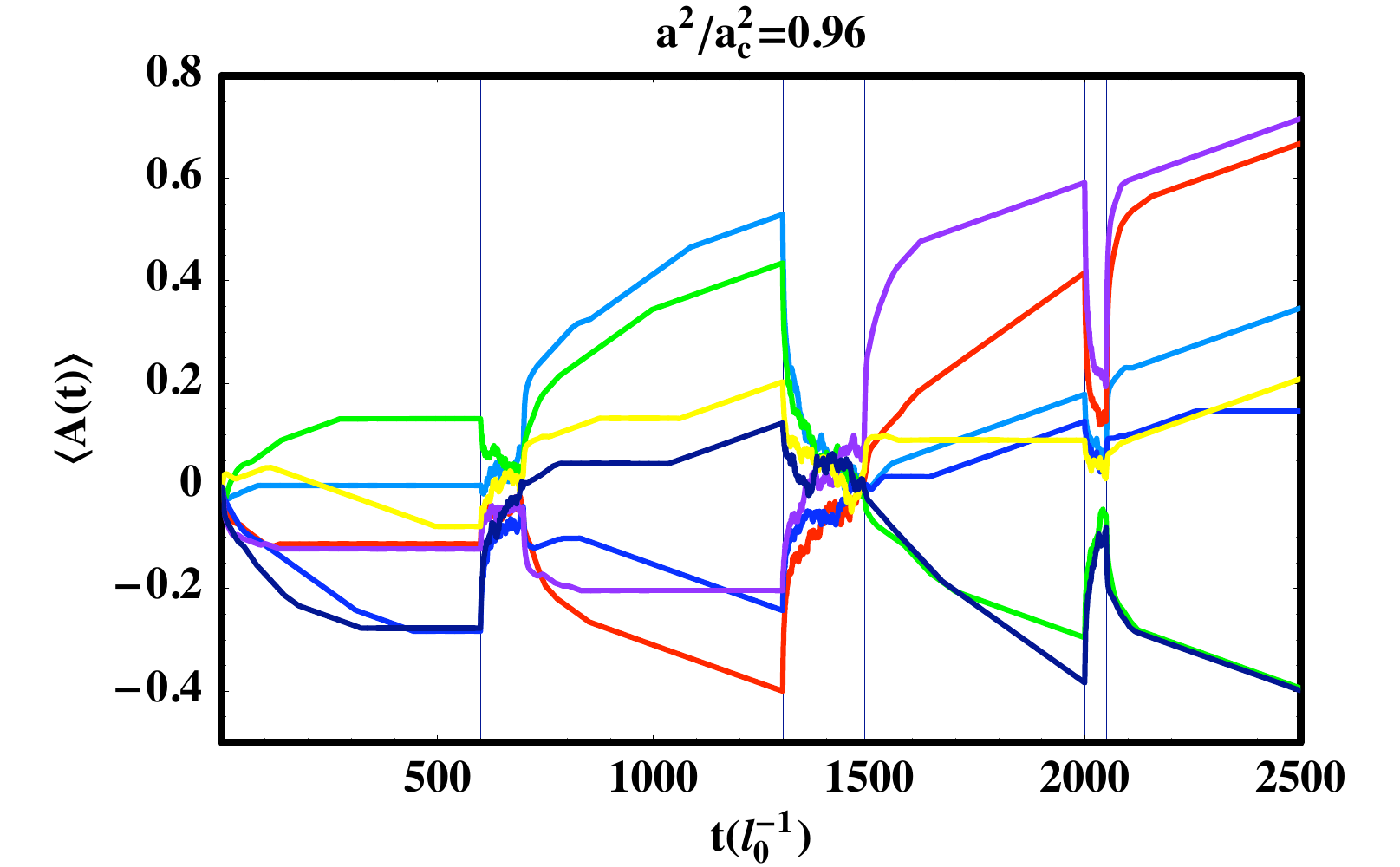}}
\caption{Punctuated Chirality. Impact of environmental effects of varying duration and fixed magnitude ($a^2/a_c^2 = 0.96$) on the evolution of prebiotic chirality in $2d$. Short events (last from left), which have little to no effect, should be contrasted with longer ones, which can drive the chirality towards purity and/or reverse its trend. (See, {\em e.g.} the green line.)} \label{fig:3event}
\end{figure}

The numerical implementation used a finite-difference leapfrog method with temporal step $\delta t = 0.005$ and spatial step $\delta x = 0.2$. Periodic boundary conditions were adopted on a $1024^2$ ($256^3$) squared (cubic) lattice in $2d$ ($3d$). Near-racemic initial conditions were prepared in two steps. First, with ${\cal S} =$1, the equation of motion for the net chirality ${\cal A}(t,x,y,z)$, eq. \ref{spatialeqns}, was solved with a simple quadratic potential, $V({\cal A})=(1/2){\cal A}^2$ and low noise ($a^2 = 0.1$) until the volumed-averaged chirality (${\cal A}(t) = (1/V)\int {\cal A}(t,x,y,z) dV$) achieved a steady-state Gaussian distribution centered in the racemic state. By steady-state we mean that $d \langle {\cal A} (t)\rangle /dt \approx 0$. Then, the noise ($w(t,x,y,z)$) was turned off  and the potential was switched to that of eq. \ref{spatialeqns}. ${\cal A}(t,x,y,z)$ relaxed deterministically toward its dynamical equilibrium state, forming domains as illustrated in Fig. \ref{fig:2Dchiral}. Of all such configurations, only those allowing for domain coexistence with near-racemic values were selected, about 38\% of our total sample. That is, we discarded all cases where ${\cal A}(t) \rightarrow \pm 1$ quickly, as these would not be sensitive to low noise environmental disturbances. The final 100 ensemble elements behaved typically as illustrated in Fig. \ref{fig:3event}: a coexisting domain structure was formed, achieving a near-static configuration for net chiralities not too far from the racemic state. At this point, the noise was turned on (at $t > 600$ in Fig. \ref{fig:3event}) in the equation for ${\cal A}(t,x,y)$, simulating the stochastic impact of environmental influences with different intensity and duration. 

\section{Results} \label{results}

In Fig. \ref{fig:3event}, we show several $2d$ runs where the environmental effects vary in duration, while their magnitude was set at $a^2/a_c^2 = 0.96$. Each colored line represents a prebiotic scenario, with sequentially occurring environmental disturbances of different duration. In order to investigate the impact of environmental effects on chiral selectivity, the scenarios reflect situations where there is no chiral selection, that is, where the two phases may coexist in dynamical equilibrium (mathematically, when $d \langle {\cal A}(t) \rangle /dt \rightarrow 0$ for ${\cal A}(t) \neq \pm 1$; chemically, in a steady state). We observe that long disturbances can drive the net chirality towards purity ($ \langle {\cal A}(t) \rangle \rightarrow  \pm 1$ for large $t$). Furthermore, note that subsequent events may erase any previous chiral bias, favoring the opposite handedness. In other words, environmental effects of sufficient intensity and duration can reset the chiral bias. This is true even if the system evolved toward homochirality prior to any environmental event.

\begin{figure}
\centerline{\includegraphics[width=5in,height=3.5in]{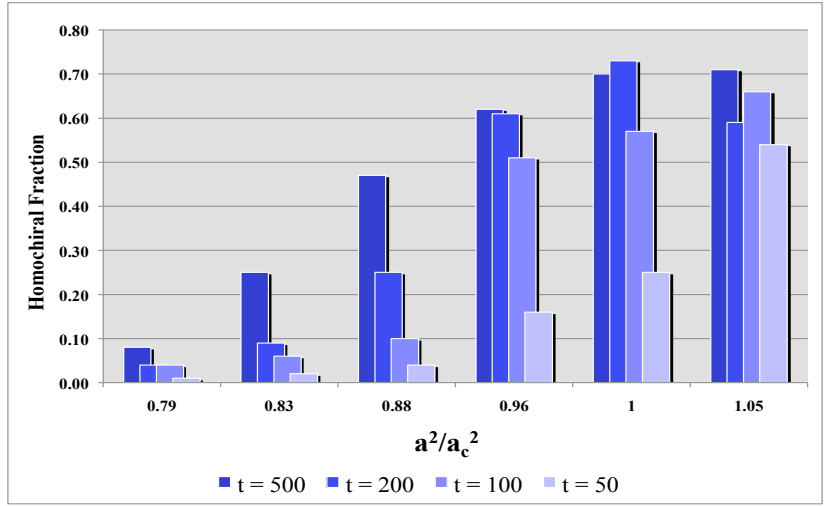}}
\caption{Fraction of $2d$ homochiral worlds after a single environmental event. Colors indicate duration of the event, while the horizontal axis labels its magnitude in units of $a_c^2$, the critical value for chiral symmetry restoration. The statistical sample included $100$ runs and the time scales of the events are in units $t(l_0^{-1})$.} \label{fig:data}
\end{figure}

In Fig. \ref{fig:data}, we summarize the results of a detailed statistical analysis of $100$ $2d$ runs that led to initial domain coexistence, that is, $d \langle {\cal A} \rangle /dt \approx 0$ (see Fig.  \ref{fig:3event} for $t < 600$). The horizontal axis displays the magnitude of the disturbance in units of $a_c^2$. The vertical axis gives the fraction of homochiral worlds, that is, those that after the disturbance obtain chiral purity. The colors represent the duration of the event. For $a^2 \geq 0.96 a_c^2$, that is, near the critical region, all but the shortest events ($t \leq 50 l_0^{-1} \approx 1.5$ months, for the nominal value of $l_0=\sqrt{2} \times 10^{-5}$s$^{-1}$ mentioned previously) lead to statistically significant chiral biasing. Results in $3d$ are qualitatively very similar, although due to heavy CPU demand we limited our analysis to 50 short runs.

It is interesting to note that the cases discarded from our statistical studies (see section \ref{num}), {\em i.e.} those that spontaneously evolve toward homochirality from near-racemic initial conditions (${\cal A}(t) \rightarrow \pm 1$ quickly) represent 62\% of the total data set. Comparing to the fraction of systems evolving toward ${\cal A}(t) \rightarrow \pm 1$ after an environmental disturbance (see Fig. \ref{fig:data}), we see that long and high-intensity events are more efficient at evolving the system toward homochirality than systems starting from an initially random near-racemic state. In other words, large disturbances can not only drive a near-racemic system toward homochirality but also can reverse the chirality of homochiral solutions, effectively erasing any previous chiral signature (see, {\em e.g.}, green line in Fig. \ref{fig:3event}). 

These results suggest, in the one extreme, that the early Earth may have played host to numerous abiogenetic events, only one of which ultimately led to the Last Universal Common Ancestor through the usual processes of Darwinian evolution. This is consistent with work indicating that life may have become globally extinct more than once \cite{Wilde}.  In the other extreme, one may consider, at the very least, that biological precursors certainly interacted with the primordial environment and may have had their chirality reset multiple times before homochiral life first evolved. In this case, our results show that separate domains of molecular assemblies with randomly set chirality may have reacted in different ways to environmental disturbances. A final, Earth-wide homochiral prebiotic chemistry would have been the result of multiple interactions between neighboring chiral domains under mechanisms already described elsewhere \cite{G, GW, BM}. 

It is interesting to compare this scenario with the suggestion that life originated in hydrothermal vents \cite{Corliss81}. One could argue that spatially separate vents consist of independent ecosystems, each thus capable of evolving life with its own chirality. And yet, what is seen today is a dominance of the usual situation with L-amino acids and D-sugars. According to our results, if life originated in distinct hydrothermal vents, it would indeed have been possible for its precursors to have evolved into distinct chirality. One would then have life of different chirality in different hydrothermal vents. Does this offer a counterexample to our scenario? Not at all. Firstly, there are several obstacles for the prebiotic synthesis of organic compounds or polymers in hydrothermal vents \cite{Lazcano_Miller}. Second, even granting the unlikely possibility that life could have originated several times over in distinct hydrothermal vents where temperatures reach $350^0$C, one must keep in mind that the whole ocean passes through them every ten million years. What happened in the distant past does not stay frozen in time. There is extensive mixing and recombination, leading to the exposure of separate chiral domains to each other and thus to competition and final preponderance of one handedness through a prebiotic version of Darwinian evolution through natural selection.

\section{Conclusions}

The possibility that environmental effects determined the homochirality of terrestrial life leads to several significant consequences. Any initial bias from extraterrestrial CPL, even if effective in the short term, would have been erased by subsequent environmental disturbances. Our results suggest that the detailed evolution of prebiotic chiral reaction networks is enmeshed with Earth's environmental history. Prebiotic Earth was a highly unstable medium, subject to active volcanism and meteoritic bombardment. Although the present analysis must be extended to describe such events in more quantitative detail, it is clear that prebiotic chirality, just as with the evolutionary history of living organisms, responded to environmental changes, alternating periods of stasis and violent activity. Our analysis predicts that other planetary platforms in this solar system and elsewhere could have developed an opposite chiral bias. As a consequence, a statistically large sampling of extraterrestrial stereochemistry would be necessarily racemic on average.

\vspace{0.5in}
The authors were partially supported by a National Science Foundation grant PHY-0653341. We had access to the NCSA Teragrid cluster under grant number PHY-070021.

\end{document}